\pdfoutput=1

\documentclass[11pt]{article}
\usepackage{graphicx}
\usepackage[final]{ACL2023}
\usepackage{float}
\usepackage{times}
\usepackage{latexsym}

\usepackage[T1]{fontenc}

\usepackage[utf8]{inputenc}

\usepackage{microtype}

\usepackage{inconsolata}

\usepackage{algorithm}
\usepackage{algpseudocode}
\usepackage{tabularx}
\usepackage{booktabs}

\usepackage{xcolor}
\definecolor{light-gray}{gray}{0.95}

\title{Building Retrieval Systems for the ClueWeb22-B Corpus}

\author{Harshit Mehrotra \\ Carnegie Mellon University \\ hmehrotr@andrew.cmu.edu \And
Jamie Callan\\Carnegie Mellon University \\callan@cs.cmu.edu \And
Zhen Fan \\ Carnegie Mellon University \\zhenfan@andrew.cmu.edu}

\begin{document}
\maketitle
\begin{abstract}
The ClueWeb22 dataset containing nearly 10 billion documents was released in 2022 to support academic and industry research. The goal of this project was to build retrieval baselines for the English section of the "super head" part (category B) of this dataset. These baselines can then be used by the research community to compare their systems and also to generate data to train/evaluate new retrieval and ranking algorithms. The report covers sparse and dense first stage retrievals as well as neural rerankers that were implemented for this dataset. These systems are available as a service on a Carnegie Mellon University cluster.
\end{abstract}

\section{Introduction}
Large scale web corpora have been essential for information retrieval research and development. The ClueWeb line of datasets is a series of such corpora. The latest in this line was released in 2022 and is called ClueWeb22 \cite{cw22}. The dataset has around 10 billion documents segmented into three categories - A, B, and L - according to the part of the web they come from. Information like raw HTML, clean text, inlink data, etc. are available for these documents in the corpus.

Information retrieval techniques like BM25 \cite{bm25}, BERT-based reranking \cite{bertrr}, and dense retrieval \cite{dpr} have been extensively studied over the last few years. Developing new retrieval algorithms or models requires comparison with such established techniques as baselines. This project was aimed at establishing such strong baselines for a section of the ClueWeb22 dataset.

Being a diverse and comprehensive dataset, such baselines would be an effective benchmark for researchers to compare their systems against. In addition to this, strong baseline systems can generate training data to develop retrieval or ranking models. Acquiring large amounts of labeled data is ideal for training deep learning retrieval models like dense encoders or rerankers, but can practically be an expensive process. Publicly available strong baseline systems like the ones we aim to build can be used as a proxy for annotators to get training data.

The project was divided into three main parts. To begin with, we created Lucene indexes to be able to implement BM25 retrieval and host it as a service. This is covered in Section \ref{sec:bm25}. After this, we explored dense retrieval where we used a pre-trained encoder to create and search on a vector index. Section \ref{sec:dense} describes this work. These two parts were focused on first stage retrieval. Finally, we used MS-MARCO \cite{marco} data combined with inlink data from ClueWeb22 to train reranker models to rank results from first stage retrieval. Details about this part can be found Section \ref{sec:reranker}.

We had access to a set of queries with manually labeled relevance judgements that we used to evaluate our implemented systems. This and other datasets used in this project are briefly described in Section \ref{sec:data}. There was also some work done previously at CMU to create retrieval systems for the ClueWeb22 corpus. We touch upon that in Section \ref{sec:prevwork}. Evaluation results of all our systems are provided and discussed in Section \ref{sec:results}. Finally, we cover directions for future work in Section \ref{sec:future}.

The code and documentation of this project can be found in the repository \url{https://github.com/harshit2997/cw22b_en_search}.

\section{Dataset}
\label{sec:data}
The ClueWeb22 corpus has 3 parts coming from different parts of the web - ClueWeb22-B containing the most popular "super head" pages of the web, ClueWeb22-A containing frequently visited "head" pages, and ClueWeb22-L containing mixed "head" and "tail" pages. ClueWeb22-B contains around 200M pages out of the 10B pages in the entire corpus. Out of these, around 87M are English documents. We restrict ourselves to these 87M documents for the purpose of this project. While one would like to search on all kinds of web documents, since this was one of the first efforts to establish some baselines, we started with the most popular segment and the language which has the most amount of search literature.

For evaluation, we had a set of around 3000 queries. A previously done BM25 implementation for first stage retrieval and a BERT reranker was used to present 10 results to Amazon Mechanical Turk annotators. This was the source of query relevance judgements. Queries that had just one word or no relevant documents were filtered out to get a final evaluation set of 2110 queries.

We used 500k inlink anchor texts from the dataset provided in \cite{anchor} to train rerankers. Each of these anchor texts has a relevant document which is the document to which the inlink points. This anchor data comes from a subset of ClueWeb22. We also used the MS-MARCO document ranking dataset \cite{marco} in addition to inlink data to train some rerankers. The dataset's train segment has around 367k queries with 384k positive judgements. Sampling of negatives for both these datasets is covered in Section \ref{sec:reranker}.

\section{Previous Work}
\label{sec:prevwork}
Some work in this area existed before this project and we describe that here. Previous contributors had done a BM25 implementation for ClueWeb22-B English documents in Apache Solr. They distributed the documents into 4 indexes. However, because of the nature of the implementation, the indexes were tied to the machines they were built on and could not be moved around.

There was also a BERT reranker used in the process of creating evaluation data as mentioned in Section \ref{sec:data}. The model was trained on MS-MARCO data and fine tuned on ClueWeb09 data. It was used to rank BM25 results to be annotated by Amazon Mechanical Turk workers.

\section{Lucene Indexes and BM25}
\label{sec:bm25}
The first task was to create a more flexible and portable BM25 implementation than what was described in Section \ref{sec:prevwork}. For this, we decided to build vanilla Lucene inverted indexes that can be moved around and searched on using Lucene or any of its wrappers. We indexed the URL, title, and body of the document's clean text JSON as searchable fields. Apart from these, we stored the index, URL, body string, and title string of the document in the index too so that these could be looked up from a retrieved document. Considering the large number of documents and our decision to place the indexes on SSDs for fast searching, we divided the 87M documents into 4 partitions and built individual indexes. 3 of these indexes were 216GB and the last one was 177GB in size. The ClueWeb22-B English documents are present in 47 small segments. The 4 partitions covered segments 0-11, 12-23, 24-35, and 36-46.

We then implemented search on these indexes using Pyserini \cite{pyserini} - a Python wrapper on Anserini \cite{anserini} that in turn is a Lucene wrapper. We used Pyserini as it is a popularly used information retrieval package and also provided easy integration into a Flask service.

After testing the implementation for high recall on our evaluation query set (see Section \ref{sec:results}), we set up 4 Flask services to be called by a client or reranker that can then aggregate and rank the four sets of results by BM25 score. Pyserini does not need the entire 216GB of RAM to load the index in memory so the service works with whatever RAM we allocate. We varied the allocated memory to observe its effect on search time. The search time remained largely stable at around 1.02 seconds per query at 95th percentile for 1000 results.

\section{Dense Retrieval}
\label{sec:dense}
After implementing BM25 for first stage retrieval, we explored a more semantic method of searching which led us to dense retrieval. In this, an encoder model is used to encode documents and store the feature vectors in a vector index that supports approximate nearest neighbor search. At query time, the encoder converts the query into a feature vector that is used to search on the dense index.

We used Contriever \cite{contriever} model fine-tuned on the MS-MARCO dataset which is available on Hugging Face \footnote{\url{https://huggingface.co/facebook/contriever-msmarco}} as our encoder. The encoder converts a sequence to a 768-d vector by mean pooling token embeddings to get sequence embeddings. The model has a maximum sequence length limit of 512 tokens, so we truncate documents/queries to that length.

The implementation is done using Pyserini \cite{pyserini} which under the hood creates a Faiss index. Faiss \cite{faiss} is a library for efficient similarity search on dense vectors. We used the most basic type of index called a "flat" index which stores vectors in fixed-width bytes, one after the other. At search time, each document vector is sequentially compared to the query vector in an exhaustive search. Search was also implemented using the Pyserini wrapper for Faiss.

We once again split the documents into 4 partitions just like when we built Lucene indexes and created Faiss indexes for each of these partitions. Each of these indexes were up to 70GB in size and were loaded into memory at the time of searching. After loading the index, each query took around 6.4 seconds (95th percentile, 1000 results) which might be reduced if we use other index types that have non-exhaustive search \footnote{\url{https://github.com/facebookresearch/faiss/wiki/Faiss-indexes}}.

The index creation process is memory extensive using around 40GB of RAM for each of the 47 smaller segments of the corpus. So, we built 47 indexes and then merged them into the final 4 partitions mentioned in Section \ref{sec:bm25}.

\section{Training Rerankers}
\label{sec:reranker}
After implementing the two first stage retrieval techniques, we moved on to training models to rerank the results of the first stage retrieval. These models are usually slower than first stage retrieval. So, they are used to rerank only the top few results of first stage retrieval. A reranker model takes as input the query and text of the search result. The output is a score that is used to rank the results.

We used BM25 for first stage retrieval because it works on the Lucene index hence the results also have the body text available. If we used dense retrieval, we would have to call Lucene separately to give the body text. So, to save some engineering effort, we went ahead with BM25 for first stage retrieval though it can be changed to dense retrieval.

To establish a baseline performance for rerankers, we first used the existing BERT reranker trained on MS-MARCO and ClueWeb09 data mentioned in Section \ref{sec:prevwork}. Once we had this baseline, we created a service to call the BM25 services mentioned in Section \ref{sec:bm25}, aggregate the results, rerank them using this BERT reranker, and finally return the top 10 results.

This existing reranker was not trained on any ClueWeb22 data. So, we next used inlink anchor text data of ClueWeb22 provided in \cite{anchor}. The dataset as around 4.5M anchor texts in the dataset out of which we used 500k. Each anchor text is treated as a query, and its relevant document is the document to which the inlink points. To create pairs of the query with negative documents, we took 30 BM25 negatives. The base model used was a pre-trained RoBERTa \cite{roberta} base model from Hugging Face. The model was trained for 2 epochs with a peak learning rate of $1e-5$. 10\% of the training steps were used to warm up the learning rate. Training was done on 4 GPUs using the Reranker library \cite{rerankerlib}.

In addition to the anchor texts, we also used the MS-MARCO document ranking dataset described in Section \ref{sec:data}. HDCT+BM25F \cite{hdct} rankings \footnote{\url{https://boston.lti.cs.cmu.edu/appendices/TheWebConf2020-Zhuyun-Dai/rankings/}} are used to sample negatives. 10 random documents from the top 100 documents in the ranking are used as negatives.

Using these 2 datasets and their combinations, we train the following rerankers:

\begin{enumerate}
    \item \textbf{AT:} Trained on 500k anchor texts with RoBERTa base as the base model.
    \item \textbf{Marco:} Trained on MS-MARCO data with RoBERTa base as the base model. URLs of the MS-MARCO dataset are also used while training.
    \item \textbf{AT->Marco:} Trained on MS-MARCO with AT as the base model. URLs of the MS-MARCO dataset are also used while training.
    \item \textbf{Marco->AT:} Trained on 500k anchor texts with Marco as the base model.
    \item \textbf{AT+MarcoNoURL:} Trained on 500k anchor texts and MS-MARCO data (without URL) with RoBERTa base as the base model.
    
\end{enumerate}

The anchor text data does not have URLs. Results of all these models are discussed in Section \ref{sec:results}.

\begin{table*}[ht]
\centering
\begin{tabular}{|l|l|l|}
\hline
\textbf{Retrieval Method}    & \textbf{Recall@500} & \textbf{Recall@1000} \\ \hline
BM25 (Lucene + Pyserini)     & 0.90                & 0.96                 \\ \hline
Dense retrieval (Contriever) & 0.67                & 0.74                 \\ \hline
\end{tabular}
\caption{First stage retrieval recall}
\label{tab:fsres}
\end{table*}

\section{Results and Discussion}
\label{sec:results}
\subsection{First Stage Retrieval}
The first evaluation was to make sure that our first stage retrieval implementations were correct. The criteria for judging this was that all or at least a high fraction of relevant documents are retrieved in the top 1000 results. 1000 is a reasonable number of documents to be processed by the downstream reranker. Recall measurements for both the first stage retrieval methods discussed earlier are presented in Table \ref{tab:fsres}.

We saw in Section \ref{sec:data} that our evaluation set's relevance judgements come from a BM25 ranking. So, it is expected that other first stage methods like dense retrieval will miss out on some of the BM25 relevant documents. However, the recall numbers for dense retrieval were very low. So, we inspected the top results for a few queries to see if the results retrieved by the dense retriever are relevant. In particular, we looked at results for queries that have very low recall.
\\
\hrule
\vspace{0.2cm}
\noindent\textbf{Query:} When was San Diego first settled \\
\textbf{Recall@1000:} 0 \\
\textbf{Relevant labeled document:} \\
What was a major difference in the settlement of Utah and the settlement of California? \\
A.      Utah was settled by Mormons hoping to escape religious persecution. \\
B.      Utah was settled in an effort to make the desert region's land sustainable. \\
C.      Utah was settled by eager prospectors looking to gain wealth by mining gold. \\
D.      Utah was settled as a way to strike a political balance between the East and the West. \\
Log in to add comment \\
ayfat23 \\
Statement that expressed the major difference that exist between settlement of Utah and settlement of California is C:Utah was settled by eager prospectors looking to gain wealth by mining gold.
American Indians were the set of people that can be regarded as settlers of Coastal California, they first arrived at San Diego, that is why San Diego is been referred to as First European settlement. This settlement pressure was from the eagerness of missionary so they can convert people to Christian.
However, the settlement of Utah was a settlement that comes from those that want to explore the land for resources such as gold so they can enrich themselves \\
Therefore, option C is correct. \\\\
\textbf{Dense retrieval result 1:}\\
History of San Diego \\
From Wikipedia, the free encyclopedia \\
The written (as opposed to oral) history of the San Diego, California, region began in the present state of California when Europeans first began inhabiting the San Diego Bay region. As the first area of California in which Europeans settled, San Diego has been described as "the birthplace of California." [1]\\
Explorer Juan Rodriguez Cabrillo claims to have discovered San Diego Bay in 1542, roughly 200 years before other Europeans settled the area;  in truth, Native Americans such as the Kumeyaay people had been living in the area for as long as 12,000 years prior to any European presence. [2]\\
A fort and mission were established in 1769, which gradually expanded into a settlement under first Spanish and then Mexican rule.\\\\
\textbf{Dense retrieval result 2:}\\
Downtown San Diego: a self-guided walking tour\\
In 1969, San Diego celebrated its 200th Birthday, honoring the Spanish settlement and mission established in 1769 by Spanish soldier Gaspar de Portola and Roman Catholic padre Junípero Serra. Yet the man known as the Father of San Diego was not an 18th Century Spaniard at all, but a 19th Century American named Alonzo E. Horton.\\
Horton, a San Francisco merchant, came to San Diego in 1867, almost 100 years after Portola and Serra. He immediately decided that the San Diego that had grown up around the first settlement was too far from the bay. Accordingly, he would buy land several miles south of that settlement and build a new San Diego.\\
The townspeople scoffed. Seventeen years earlier, another man, William Heath Davis, had the same idea, tried it, and gave up in defeat two years later. But Horton was undaunted. He bought 960 acres of barren land for a then-generous price of \$265 (an average of 27-½ cents per acre), and began to build his New Town. New Town had its economic ups and downs, and on one of the downs Horton himself lost almost all the money he had made, but the city persevered. The original San Diego settlement survives as Old Town, a restored reminder of Old California, but New Town is today's downtown San Diego\\
\hrule
\begin{table*}[ht]
\centering
\begin{tabular}{|l|l|l|l|l|}
\hline
\textbf{Reranker}      & \textbf{URL} & \textbf{MRR} & \textbf{P@5} & \textbf{P@10} \\ \hline
None (BM25)            & -            & 0.258        & 0.116        & 0.097         \\ \hline
Existing BERT          & Yes          & 0.330        & 0.144        & 0.110         \\ \hline
AT                     & No           & 0.078        & 0.027        & 0.023         \\ \hline
Marco                  & Yes          & 0.273             &  0.115            &   0.090            \\ \hline
AT-\textgreater{}Marco & Yes          & 0.296        & 0.126        & 0.095         \\ \hline
Marco-\textgreater{}AT & No           & 0.044        & 0.014        & 0.013         \\ \hline
AT+MarcoNoURL          & No           & 0.185        & 0.079        & 0.063         \\ \hline
\end{tabular}
\caption{Reranker MRR, P@5, P@10 measurements. Reranker names are the ones described in Section \ref{sec:reranker}}. All rerankers are applied to top 1000 BM25 results.
\label{tab:rrres}
\end{table*}
It is quite evident that the dense retrieval results directly answer the query with the first result directly being the Wikipedia page for the history of San Diego. On the other hand, the relevant labeled document is an answers forum page predominantly about the settlement of Utah with a sentence talking about American Indians landing in San Diego. That sentence too does not discuss the settlement of San Diego. This example brings out the lexical match nature of BM25 and the semantic (hence maybe more relevant) matching nature of dense retrieval.

Another example is the query "top grocery shops in United States". The 2 relevant labeled documents are about Korean grocery stores in Mississippi and a single grocery store in the city of Kalona. The dense retriever surfaces results about 10 best readers' choice supermarkets and the most popular grocery store in every state. Once again the dense retriever seems to be retrieving more relevant results.

There were a few more queries that we inspected in a similar way and concluded that the relevance judgements we have are probably too biased towards lexical matching. This means we need to annotate a larger set of results for each evaluation query and not just 10 results to be able to make a fairer comparison between BM25 and dense retrieval. Another option is to let annotators run the query and interleave the results of BM25 and dense retrieval for them to interact with.

\subsection{Reranking}
As mentioned in Section \ref{sec:reranker}, we first used the existing reranker to establish a baseline reranker performance. This reranker was trained to make use of the URL of the document apart from the title and body. Then we trained more rerankers described in the same section. After getting initial results from BM25, the goal of reranking was to push relevant results to higher ranks. Hence, we would want to improve MRR (mean reciprocal rank) and precision at higher positions (say 5, 10). A summary of reranker performance is presented in Table \ref{tab:rrres}. The column 'URL' indicates whether the URL of a document was used at inference time along with its title and body.

\begin{figure}[h]
    \centering
    \includegraphics[scale=0.55]{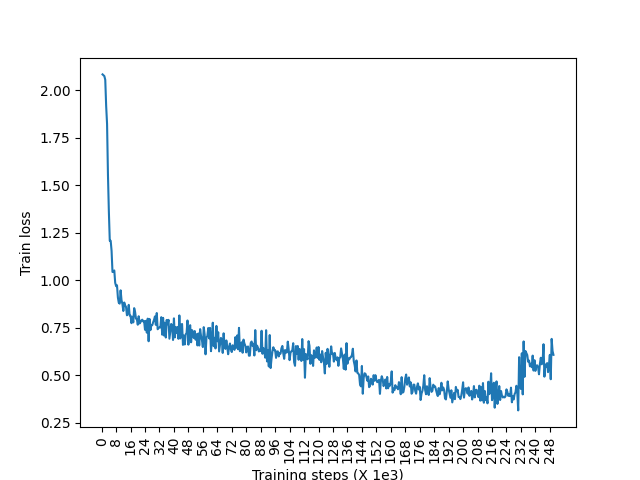}
    \caption{Train loss for the reranker trained on anchor text}
    \label{fig:loss}
\end{figure}

The first observation from these results is that whenever the anchor text is used as the final stage of training (AT and Marco->AT), the performance degrades severely - even worse than the first stage retrieval. We looked at the training loss graph of the better of these two, i.e., AT to see if we can take models at checkpoints with different losses and if that makes a difference. The training loss is shown in Figure \ref{fig:loss} and the MRR at 3 checkpoints other than the final one (250k steps) are shown in Table \ref{tab:chkpt}.

\begin{table}[h]
\centering
\begin{tabular}{|c|c|c|}
\hline
\textbf{Training Step} & \textbf{Loss} & \textbf{MRR} \\ \hline
32k                    & 0.752         & 0.167        \\ \hline
176k                   & 0.440         & 0.117        \\ \hline
216k                   & 0.370         & 0.119        \\ \hline
250k                   & 0.608         & 0.078        \\ \hline
\end{tabular}
\caption{MRR of the AT reranker at different checkpoints}
\label{tab:chkpt}
\end{table}

While a relationship between loss and MRR is not clear, it does seem like training more with anchor text degrades performance. This is also supported by the observation that Marco performs better than when the model is trained again using anchor text data, i.e., in Marco->AT.

We also see that AT->Marco which was trained on MS-MARCO with the anchor text model as the starting point does much better than AT and even improves performance over BM25. The second stage training with MS-MARCO was able to fix the incorrect behavior learnt from anchor text data to a great extent.

An interesting observation is that AT->Marco does better than just Marco. So, maybe there is some benefit of first training with anchor text data that needs to be investigated.

Finally we see that AT+MarcoNoURL which uses a mix of MS-MARCO and anchor text data for training gives performance that is in between rerankers that use just anchor text for final training (AT, Marco->AT) and rerankers that use just MS-MARCO for final training (AT->Marco, Marco). This is another indication towards the detrimental nature of anchor text data for final training and the beneficial nature of MS-MARCO data.

To understand the effect of anchor text data on the reranker, we took the best AT reranker checkpoint out of the ones we checked earlier, i.e., the one at 32k training steps and looked at retrieval results for a few queries just like we did for the dense retriever. We compared this reranker with the old BERT reranker to see what kind of results are promoted by both rerankers. Certain excerpts from results are shown below.
\\
\hrule
\vspace{0.2cm}
\noindent\textbf{Query:} Who can introduce new bills\\
\textbf{BERT result 1:}\\
Act of Parliament From Wikipedia, the free encyclopedia\\
Acts of parliament, sometimes referred to as primary legislation, are texts of law passed by the legislative body of a jurisdiction (often a parliament or council ). [1] In most countries with a parliamentary system of government, acts of parliament begin as a bill, which the legislature votes on. Depending on the structure of government, this text may then be subject to assent or approval from the executive branch \\
Bills[edit]\\
A draft act of parliament is known as a bill. In other words, a bill is a proposed law that needs to be discussed in the parliament before it can become a law. In territories with a Westminster system, most bills that have any possibility of becoming law are introduced into parliament by the government. This will usually happen following the publication of a "white paper", setting out the issues and the way in which the proposed new law is intended to deal with them. A bill may also be introduced into parliament without formal government backing; this is known as a private member's bill . In territories with a multicameral parliament, most bills may be first introduced in any chamber. However, certain types of legislation are required, either by constitutional convention or by law, to be introduced into a specific chamber....\\\\
\textbf{BERT result 2:}\\
How Laws Are Made  | USAGov\\
....\\
Congress has two legislative bodies or chambers: the U.S. Senate and the U.S. House of Representatives. Anyone elected to either body can propose a new law. A bill is a proposal for a new law.\\
Steps in Making a Law\\
A bill can be introduced in either chamber of Congress by a senator or representative who sponsors it. Once a bill is introduced, it is assigned to a committee whose members will research, discuss, and make changes to the bill. The bill is then put before that chamber to be voted on. If the bill passes one body of Congress, it goes to the other body to go through a similar process of research, discussion, changes, and voting. Once both bodies vote to accept a bill, they must work out any differences between the two versions.....\\\\
\textbf{AT (32k) result 1:}\\
All of the congressional proposals to change Section 230.\\
All the Ways Congress Wants to Change Section 230\\Republicans and Democrats alike want to change Section 230 of the Communications Decency Act.\\
....\\
As Section 230 became a fixture of election debates and Trump’s Twitter feed, legislators from both sides of the aisle introduced bills to try to address their concerns. Democrats introduced bills that reduced the scope of Section 230 protections in civil rights cases and terrorism cases, while Republicans introduced bills that sought to compel platforms to be more “neutral” in their moderation of online content. And there were even a few bipartisan proposals, which focused on child sexual exploitation, content moderation operations, and reduced protections for content that courts determine to be illegal. Some of the proposals may be reasonable; others could collide with the First Amendment. With a flurry of bills introduced in 2020 and 2021—and roughly 12 bills introduced in the last four months of 2020 alone—it’s been tough for researchers, company employees, and policymakers to keep track....\\\\
\textbf{AT (32k) result 2:}\\
The UK Parliament; Public policy engagement toolkit; Newcastle University\\
....\\
Both Houses must agree on a Bill before it can become law, although the Lords has a limited say on financial bills. The Lords can delay Bills and make the Commons to reconsider its decisions. In this capacity, the Lords acts as a check on the House of Commons that is independent from the electoral process. The Lords can also introduce Bills, although this is less common. Unlike the Commons, there is no fixed limit on the number of Members of the House of Lords (currently, there are around 825).\\
....\\
A Green Paper is often the first step towards introducing a Bill into Parliament. Following the publication and discussion of the Green Paper, the department may release a White Paper, which will also be published on departmental websites. This is a more detailed and formalised version of the Green Paper, and often forms the basis for a Bill to be introduced to Parliament. If the government believes there is an urgent need for particular legislation, or wishes to limit opportunities for consultation, the department may publish a White Paper first, without publishing a Green Paper. Government is not obliged to produce Green or White Papers, and in some cases a Bill may be introduced to Parliament without any public consultation steps.\\
Parliamentary stages of a Bill\\
Bills can be introduced into either the Commons or the Lords, and each Bill goes through set stages of legislation in each House. The Bill must be approved by both Houses before it receives Royal Assent, after which it becomes an Act of Parliament.\\
\hrule \vspace{0.2cm}
We see that the BERT reranker's results are more relevant to the query with the first one providing a general global mechanism for introducing a bill and the second one answering the question for the US. However, the AT reranker's first result is an article about certain specific bills that were introduced recently. The second result talks about the process of passing a bill in the UK but does not mention who introduces a bill. We inspected a few more queries, and it seems like the anchor text-trained reranker promotes results tangentially related to the query. This can be an artifact of the data since anchor texts are generally related to the current page but don't exactly capture the contents of a page. However, it is interesting that the authors of \cite{anchor} were able to train a good dense retriever using this data. 

\section{Conclusion and Future Work}
\label{sec:future}
We investigated the performance of different search techniques for English ClueWeb22-B documents in this project. We implemented BM25 and dense retrieval for first stage retrieval, and reranked the results using rerankers. The best reranker we trained was first trained on anchor text data followed by MS-MARCO document ranking data. However, the existing BERT reranker is still better than the one we trained. Finally, we hosted all these retrieval methods as APIs on a CMU cluster.

Going ahead there are a few directions to be explored further. Qualitative analysis showed the dense retriever to have better results than the annotated relevant ones. Better quality relevance data or interleaved judgements need to be gathered to ascertain if BM25 or dense is the better first stage retrieval. There is also scope to improve the latency of dense retrieval by exploring alternate index types in Faiss.

On the reranker front, more investigation is required into the anchor text data - specifically the effect of the size of data, effectiveness as first stage training versus standalone training. Ablations should also be conducted to find a better hyperparameter setting for training. Finally, we should also understand why anchor text data is good for training dense retrievers but not rerankers.

\bibliographystyle{acl_natbib}

\end{document}